\documentclass{aa}  

\usepackage{longtable} 
\usepackage{booktabs}
\usepackage{graphicx}
\usepackage{txfonts}
\usepackage{hyperref}
\makeatletter
\let\do@mlinenumbers\relax
\let\linenumbers\relax
\let\endlinenumbers\relax
\let\linenumbersep\relax
\let\thelinenumber\relax
\let\makeLineNumber\relax
\makeatother
\begin{document}

   \title{A homogeneous TTV investigation of all TESS systems with a confirmed single transiting planet}

   \author{Luca Naponiello\inst{1}}

   \institute{$^{1}$INAF -- Osservatorio Astrofisico di Torino, Via Osservatorio 20, 10025 Pino Torinese, Italy\\
              \email{luca.naponiello@inaf.it}
             }

   \date{Received: October 01, 2025; Accepted: November 20, 2025.}

 
  \abstract
   {Transit Timing Variations (TTVs) are a powerful tool for detecting unseen companions in systems with known transiting exoplanets and for characterizing their masses and orbital properties. Large-scale and homogeneous TTV analyses are a valuable method to complement the demographics of planetary systems and understand the role of dynamical interactions.}
   {We present the results of a systematic TTV analysis of 423 systems covering $\sim16\,000$ transits, each with a single transiting planet first discovered by the NASA TESS mission and then confirmed or validated by follow-up studies. The primary aim of this survey is to identify the most promising candidates for dynamically active systems that warrant further investigation.}
   {Our analysis is performed in a two-stage pipeline. In the first stage, precise measurements of individual transit times are extracted from the TESS light curves for each system in an homogeneous way. 
   In the second stage, we apply a two-tiered decision framework to classify candidates by analyzing the resulting transit variations. 
   Based on excess timing scatter ($\chi^2_{\mathrm{mod}}$) and the difference in Bayesian Information Criterion ($\Delta$BIC) of periodic models over linear ones, the TTVs are classified as significant, marginal, or non-detections.}
   {We find 11 systems with significant TTVs, 5 of which were announced in previous works, and 10 more systems with marginal evidence in our sample. We present three-panel diagnostic plots for all the candidates, showing phase-folded light curves, the transit variations over time, and the same variations folded on the recovered TTV period. A comprehensive summary table detailing the fitted parameters and TTV significance for the entire survey sample is also provided.}
   {This survey constitutes the largest homogeneous TTV analysis of TESS systems to date. 
   We provide the community with updated ephemerides and a catalogue of high-quality TTV candidates, enabling targeted follow-up observations and detailed dynamical modelling to uncover the nature of unseen companions and study system architectures.}


   \maketitle
%

\section{Introduction}

The transit method, particularly as employed by space missions like Kepler \citep{Borucki2010} and TESS \citep{Ricker2015}, has revolutionized exoplanet science. Beyond planet discovery, high-precision photometry enables the study of subtle dynamical effects, the most prominent of which is the Transit Timing Variation (TTV). Gravitational perturbations from companion planets or other stars, as well as effects such as orbital decay (see, e.g., \citealt{Yee2020}), can cause transits to deviate from the times predicted by a simple linear ephemeris \citep{Holman2005,Agol2005}. In some systems, both tidal decay and additional companions can contribute to the observed TTVs \citep{Harre2024,Alvarado2025}. The detection and characterization of TTVs has therefore the potential to reveal the masses of interacting planets and constrain their orbital properties, often providing insights into system architectures that are inaccessible by other means (e.g. \citealt{Hadden2017,Agol2021}).

The TESS mission is observing the brightest stars across the entire sky, yielding an ever-growing sample of confirmed transiting exoplanets \citep{Kunimoto2022}. With the $\sim$7 year temporal baseline provided so far by TESS observations, conditions are now favourable for a comprehensive TTV analysis, offering an unprecedented opportunity to conduct large-scale surveys to measure the frequency and amplitude of dynamical interactions across a wide variety of systems, similarly to what has previously been done for Kepler (e.g. \citealt{Holczer2016}). Recent TTV studies using TESS data have mainly targeted hot Jupiters (e.g. \citealt{Ivshina2022,Wang2024}), providing early insights into the detectability of dynamical interactions and the current limitations of TTV analyses. Large-scale surveys of this kind are critical to building a statistical picture of planet formation and evolution \citep{Mazeh2013,Hadden2017}, but have so far been carried out only for Kepler systems; similar initiatives have been encouraged for TESS \citep{Yalahomi2024}.

Here, we present the results of a large and homogeneous TTV survey of 423 systems with a single confirmed transiting planet from the TESS catalogue. In this context, \textit{homogeneous} refers to methodological uniformity rather than astrophysical similarity among targets. This approach ensures that all TTV detections and non-detections are directly comparable, as they result from a single, self-consistent analysis pipeline applying identical priors, statistical thresholds, and vetting procedures. By focusing on single-transiting systems, we search for additional companions that do not transit, either due to unfavourable orbital inclination, small size, or long orbital periods beyond the current observational baseline, but that may still be in orbital resonance with the transiting planet or massive and eccentric enough to induce detectable timing variations. This approach can help constrain the true occurrence rate of multi-planet systems (e.g. \citealt{Xie2014}). 

We describe the analysis pipeline, from initial data processing to the final, rigorous classification of TTV candidates, respectively in in Sections\,\ref{sec:pipeline} and \ref{sec:vetting}. Then, we present our results, including our significant TTV candidates, in Sect.\,\ref{sec:results}, and finally draw the conclusions of the work in Sect.\,\ref{sec:conclusions}.


\section{Data and analysis pipeline}
\label{sec:pipeline}

\subsection{Target sample and light curves}
Our starting sample consists of systems listed in the TESS Objects of Interest (TOI; \citealt{Guerrero2021}) catalogue on the Exoplanet Follow-up Observing Program (\href{https://exofop.ipac.caltech.edu/tess/}{ExoFOP-TESS}) webpage, restricted to those with a single confirmed planet (``CP'' disposition\footnote{This should not be confused with the known planet (KP) disposition, which designates planets identified prior to TESS by other facilities. In this work, we assume that the CP transit dataset, unlike KP, is primarily composed of TESS observations.}, at least up until September 2025). We exclude mono-transiting planets and a small number of particularly complex planets from the sample (about 40 systems in total), such as LHS 3844\,b (TOI-136.01; \citealt{Vanderspek2019}), a super-Earth with >200 transits observed by TESS to this day, due to computational constraints. Specifically, targets that largely exceeded five days of processing time on our computing system are omitted.

For each of the remaining 423 targets, we use the Presearch Data Conditioning Simple Aperture Photometry (PDC-SAP; \citealt{Stumpe2012,Stumpe2014,Smith2012}) light curves as taken from the Mikulski Archive for Space Telescopes (MAST) archive, and processed by the TESS Science Processing Operations Center (SPOC; \citealt{Jenkins2016}) pipeline\footnote{We did not use the Quick-Look Pipeline (QLP; \citealt{Fausnaugh2018}) light curves since SPOC products were available in most cases, and QLP data are generally noisier and less reliable (see, e.g, appendix A of \citealt{Fernandes2025}). A proper use of QLP would require careful, system-by-system reduction, which was not appropriate for the automated, survey-wide approach adopted here.} 
at the NASA Ames Research Center, employing the \texttt{lightkurve} package \citep{lightkurve2018}. In particular, the PDC-SAP light curves are already corrected for long-term modulations and instrumental systematics \citep{Jenkins2016}. When multiple cadences are available, the fastest one is always preferred (down to 2-minutes\footnote{For some systems the 20-second cadence was also available, though the difference in transit time precision is often negligible, in comparison with the 2-minute cadence, while the increase in computational time is substantial.}). 

The light curves are then prepared for the analysis. This preparation involved a cleaning step where any data point with invalid flux values or errors is removed. In some extreme cases, entire sectors are excluded as they are made entirely of negative flux values (e.g. Sector 37 for TOI-1231, which is likely due to bad smear correction issues due to bright stars in the focal plane). Light curves from different TESS sectors are then stitched together to form a single, long-baseline time series for each cadence. For sectors observed at longer cadence, the transit models are supersampled (i.e. computed at higher temporal resolution and then averaged over each exposure) in order to match the effective cadence of the short-cadence data and ensure consistent fitting. This is implemented within the \texttt{juliet} framework \citep{Espinoza2019}, which uses the \texttt{batman} package \citep{Kreidberg2015} to generate the transit models with built-in supersampling capabilities. Then, we identify a time window ($\tau$) around each expected transit and remove data-points that fall outside. The window is usually defined as:
\begin{equation}
\tau \leq |\,T_0-2d\,|
\end{equation}
where $d$ is the duration of the transit. This step mainly reduces computation time while ensuring that the fitted region contains enough out-of-transit data to model the local baseline. In some cases, we employ larger windows because of extensive TTVs (e.g. TOI-139.01) or severe stellar activity (e.g. TOI-1228), improving the fit to the local light-curve modulation. Nevertheless, since only short segments around each transit are analyzed, the broader stellar variability is not fully captured. When the star is very active, a global fit to the full light curve may yield more precise transit timings, though at a higher computational cost. We visually inspected the most challenging cases (such as TOI-1228) and verified that, after increasing the window size, the transits remain well captured and properly modelled. Finally, transits without at least 60\% coverage are also removed as their centre times are mostly uncertain.

\subsection{Transit timing measurements}
The precise time of each transit is measured using the \texttt{juliet} package. We employ a Bayesian framework coupling the transit model with a Gaussian Process (GP) via an approximate Matern kernel ($\epsilon=0.01$) to account for time-correlated noise (this kernel converges to the Matern-3/2 form as $\epsilon\to0$; \citealt{Foreman2018,Espinoza2019}).

Instead of fitting a linear ephemeris, we treat the centre time of each individual transit $(T_n)$ as a free parameter in the model. Priors for these times are set as Normal distributions centred on the expected time from the catalogue ephemeris, with a standard deviation large enough to allow for significant variations ($\tau/5$). The priors are set to normal distributions also for the stellar density $(\rho_{\star})$, planet-to-star radius ratio $(R_p/R_{\star})$, impact parameter $(b)$, eccentricity and argument of periastron $(e,\omega)$\footnote{Eccentricity is allowed to vary only when $e>0.1$; otherwise, we fix $e=0$ and $\omega=90^\circ$. For systems with truly small eccentricities, the transit shape is still accurately modelled, although the inferred stellar density may differ slightly from the real value to properly account for the transit duration.} based on a query to the NASA Exoplanet Archive \citep{Christiansen2025}, starting from the reference automatically chosen by the database. Finally, the model is fit\footnote{The analyses have been performed on an HPE ProLiant DL560 Gen10 rack server equipped with 2$\times$ Intel Xeon Gold 6252N processors (2.3\,GHz, 24 cores, 150\,W each) and 128\,GB of RAM, hosted at INAF - Osservatorio Astrofisico di Torino.} using the \texttt{dynesty} nested sampler \citep{Speagle2020} which, unlike traditional Markov Chain Monte Carlo (MCMC) algorithms, simultaneously estimates the posterior distributions and the Bayesian evidence, allowing for a more efficient exploration of multimodal or highly correlated parameter spaces \citep{Skilling2004}.

\section{Candidate vetting and identification of TTVs}
\label{sec:vetting}
The observed minus calculated (O$-$C) transit times generated in the first stage are passed to a second analysis pipeline designed to identify periodic signals and classify their significance.

\subsection{Outlier rejection}
Before searching for periodicity, we apply two filters to the O$-$C data to ensure robustness against outliers. First, we discard any transit measurement whose error bars are excessively large (greater than 3 times the mean error). Second, we perform a 5-$\sigma$ clipping on the O$-$C values themselves to remove any extreme timing outliers. Transits from sectors with long exposure times were in some cases (e.g. TOI-1516.01 and TOI-2046.01) considerably noisier\footnote{Longer exposures reduce temporal resolution, smoothing the ingress and egress and thus limiting the precision of the derived transit parameters. This effect is more pronounced for short-duration transits, while longer events remain well sampled and often benefit from the improved photometric stability of long-cadence data.} than the rest and were thus excluded.

\subsection{Periodicity search}
We use the \texttt{astropy.timeseries.LombScargle} class \citep{Robitaille2013} to compute a frequency periodogram of the filtered O$-$C data, evaluated over the range $f_{\rm min} \sim 1/(10 \Delta T)$ to $f_{\rm max} \sim 1/(2 P_{\rm orb})$, where $\Delta T$ is the total baseline and $P_{\rm orb}$ the orbital period. The periodicity search is only performed if a system has at least 3 valid transit measurements. The upper frequency limit corresponds to the effective Nyquist frequency of the transit sampling, ensuring that the periodogram includes all physically meaningful TTV signals within the data resolution. This naturally imposes a “Nyquist floor” for TTVs, below which true variations cannot be reliably recovered; periods shorter than twice the orbital period would instead appear as aliases of the real modulation \citep{Yalahomi2024}. Within the physically defined bounds $f_{\rm min}$ and $f_{\rm max}$, the detection efficiency mainly depends on the timing precision and the number of observed transits. 

\begin{figure*}
\centering
\includegraphics[width=0.8\textwidth]{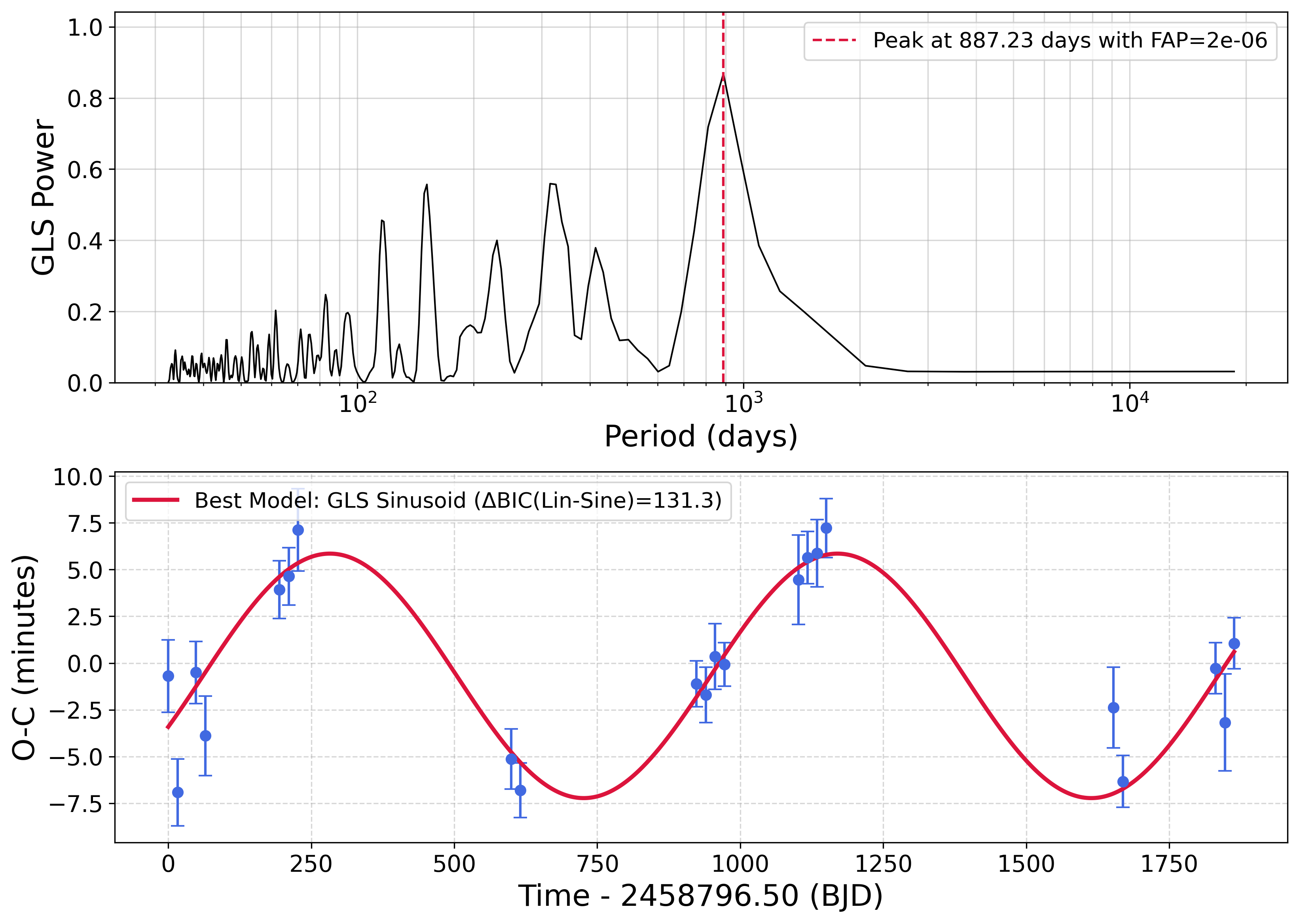}
\caption{\textit{Top panel}: Generalized Lomb–Scargle (GLS) periodogram of the TOI-1611\,b O$-$C transit times, with the best periodicity highlighted by a red vertical line. \textit{Bottom panel}: O$-$C measurements over time, with the best-fitting sinusoidal model shown in red.}
\label{fig:GLS}
\end{figure*}

Fig.\,\ref{fig:GLS} shows a sample periodogram for TOI-1611\,b, highlighting the best periodicity and the corresponding false-alarm probability (FAP), together with the best-fitting sinusoid derived from that frequency.

\subsection{Model selection and candidate classification}\label{sec:classification}
The core of our vetting process relies on a two-tiered statistical test to determine if a periodic TTV model is justified by the data. First, we compare the preferred sinusoidal model against a simple linear model (representing no TTVs or a secular trend). We use the Lomb-Scargle algorithm to find the best-fit sinusoid at the peak frequency $f_{\mathrm{TTV}}$, considering both a single-term ($N=1$) and double-term ($N=2$) harmonic fit and selecting the one with the superior bayesian information criterion (BIC). We then compute the $\Delta$BIC between this best sinusoidal model and the linear model:
\begin{equation}
\Delta\mathrm{BIC} = \mathrm{BIC}_{\mathrm{linear}} - \mathrm{BIC}_{\mathrm{sinusoid}}.
\end{equation}
A periodic model is considered strongly preferred only if $\Delta\mathrm{BIC} \geq 10$ \citep{Kass1995}, indicating substantial evidence against a simple linear ephemeris.

Second, even if a periodic model is preferred, the signal must have a significant amplitude relative to the measurement uncertainties. We quantify this using a scatter statistic, $\chi^2_{\mathrm{mod}}$, defined as:
\begin{equation}
\chi^2_{\mathrm{mod}} = \frac{\mathrm{MAD}(\mathrm{O-C})}{\mathrm{median}(\sigma_{\mathrm{O-C}})}
\end{equation}
where MAD is the median absolute deviation, a robust estimator of the standard deviation (see, e.g., \citealt{Holczer2016}). This statistic measures the ratio of the observed scatter in the O$-$C data to the expected scatter from the median photometric timing precision, $\sigma_{\mathrm{O-C}}$. Our targets are classified based on both criteria as:

\begin{itemize}
    \item \textbf{Strong candidates.} The TTVs are significant, with both strong model preference ($\Delta\mathrm{BIC} \geq 10$) and highly significant scatter ($\chi^2_{\mathrm{mod}} \geq 3.0$).
    \item \textbf{Weak candidates.} Marginal detections, having strong model preference ($\Delta\mathrm{BIC} \geq 10$) but only moderate scatter ($2.0 \leq \chi^2_{\mathrm{mod}} < 3.0$).
    \item \textbf{Non-detections.} Cases where either the periodic model is not statistically preferred ($\Delta\mathrm{BIC}<10$) or the timing scatter is consistent with the measurement errors ($\chi^2_{\mathrm{mod}} < 2.0$).
\end{itemize}

We acknowledge that these thresholds are somewhat arbitrary and may differ in other studies; for transparency, we report the $\Delta\mathrm{BIC}$ and $\chi^2_{\rm mod}$ values for each system in Table\,\ref{tab:ttv_summary}, and the full O$-$C list online, allowing follow-up studies to apply alternative criteria or select interesting targets for further investigations. We further note that the adopted thresholds could in principle introduce a bias against low signal-to-noise or long-period signals, where the statistical significance of a TTV pattern is harder to establish. Additional validation efforts, such as injection–recovery tests aimed at quantifying the detectability of synthetic TTV signals under realistic noise conditions, would help to characterize these effects more precisely, but performing such simulations would require significant computational resources and was therefore beyond the scope of this work.


\section{Results}\label{sec:results}
The analysis of our 423-system sample yields 6 and 10 new candidates with strong and weak TTVs, respectively. For each of those, we generate a three-panel diagnostic plot to facilitate visual inspection and further study (Figs.\,\ref{fig:TTV_1}, \ref{fig:TTV_2}, and \ref{fig:TTV_3}). The panels display: (1) the phase-folded transit light curve, corrected for the measured TTVs to show the quality of the underlying transit fit, (2) the O$-$C diagram as a function of time, and (3) the O$-$C diagram phase-folded to the detected TTV period with the best-fit sinusoidal model overplotted, which highlights the periodic nature of the signal.

In addition, in Table\,\ref{tab:ttv_summary} we provide a comprehensive summary of the derived orbital and TTV parameters for every system in the survey. Other than the $\Delta$BIC and $\chi^2_{\mathrm{mod}}$ values, the table includes the TTV classification, the TTV period ($P_{\mathrm{TTV}}$) and semi-amplitude ($A_{\mathrm{TTV}}$) for candidates, and the updated parameters ($P_{\mathrm{orb}}$, $T_0$, $R_p/R_{\star}$, etc.) for all targets. In some cases, such parameters differ from published values owing to the inclusion of many new transits. For instance, the orbital period of TOI-139\,b turns out to be $\sim20$ minutes longer than the one presented in \citealt{Mistry2023}, whereas for the peculiar planet TOI-1690.01 / WD\,1856+534\,b, hosted by a white dwarf \citep{Vanderburg2020}, we find a lower impact parameter and planet-to-star radius ratio. Conversely, for targets significantly affected by flux contamination from nearby sources, the planet-to-star radius ratio may be slightly underestimated, as no additional dilution correction was applied beyond the default PDC-SAP adjustment based on the TESS Input Catalog.

The results of the fits have been visually inspected to assess both the quality of the transit modelling and the derived transit timing variations. While this inspection confirms that the fits are generally robust, obtaining extremely accurate TTV measurements is beyond the scope of this work, as it would require individualized modelling of each system. For instance, certain systems with a large number of transits could benefit from a higher number of live points in the nested sampling procedure; in this study, we limited the number of live points to 1000 or 5000 depending on the number of transits (below and above 30, respectively) to reduce computation time. Additionally, some targets in our sample have also been observed from the ground or by other space missions, and a comprehensive TTV study should incorporate all available transits (see, e.g., the analysis of TOI-1227\,b TTVs in \citealt{Almenara2024}).

This approach prioritizes a homogeneous, large-scale analysis, while leaving the detailed characterization of individual systems for follow-up studies. Furthermore, we opted not to fit independent transit durations for each transit, both to reduce computation time and because individual TESS transits generally lack the photometric precision to accurately constrain them\footnote{Typical transit duration variations observed in Kepler systems are on the order of $\sim1$ minute over several hundred days \citep{Holczer2016,Shahaf2021}, well below the average O$-$C uncertainty of $\sim5.8$ minutes in our sample. Moreover, the irregular and often widely separated sector coverage of TESS further limits the detectability of such subtle duration changes.}, unlike Kepler (see, e.g., \citealt{Boley2020,Kaye2025}). Indeed, to our knowledge, transit duration variations for TESS have been observed in very few cases (e.g. TOI-1338\,A\,b, where the variations are attributed to the misalignment between the planet's orbit and the binary stars - \citealt{Kostov2020}).

\begin{figure*}
\centering
\includegraphics[width=1\textwidth]{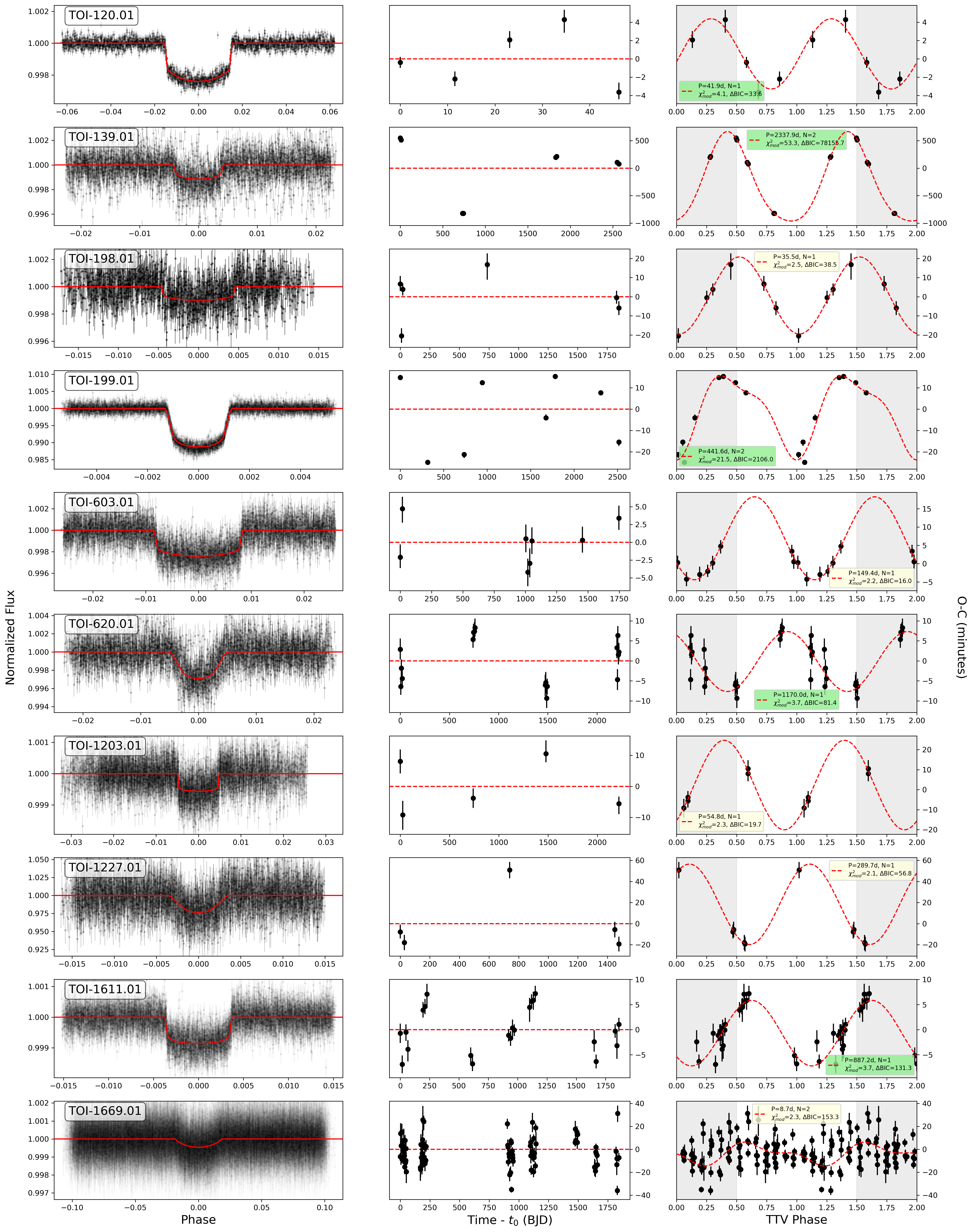}
\caption{Three-panel diagnostic plot for TTV candidates. \textit{Left:} The TTV-corrected, phase-folded transit light curve with the best-fit model. \textit{Middle:} O$-$C measurements over time from first transit ($t_0$, which is often, but not always, the same as $T_0$ from Table\,\ref{tab:ttv_summary}). \textit{Right:} The O$-$C diagram folded to the best fit TTV period (with the model represented by a dashed red line). The colour of the legend indicates the classification, with green and light yellow representing strong and weak candidates, respectively. Shaded regions indicate phase repetitions for visual continuity.}
\label{fig:TTV_1}
\end{figure*}

\begin{figure*}
\centering
\includegraphics[width=1\textwidth]{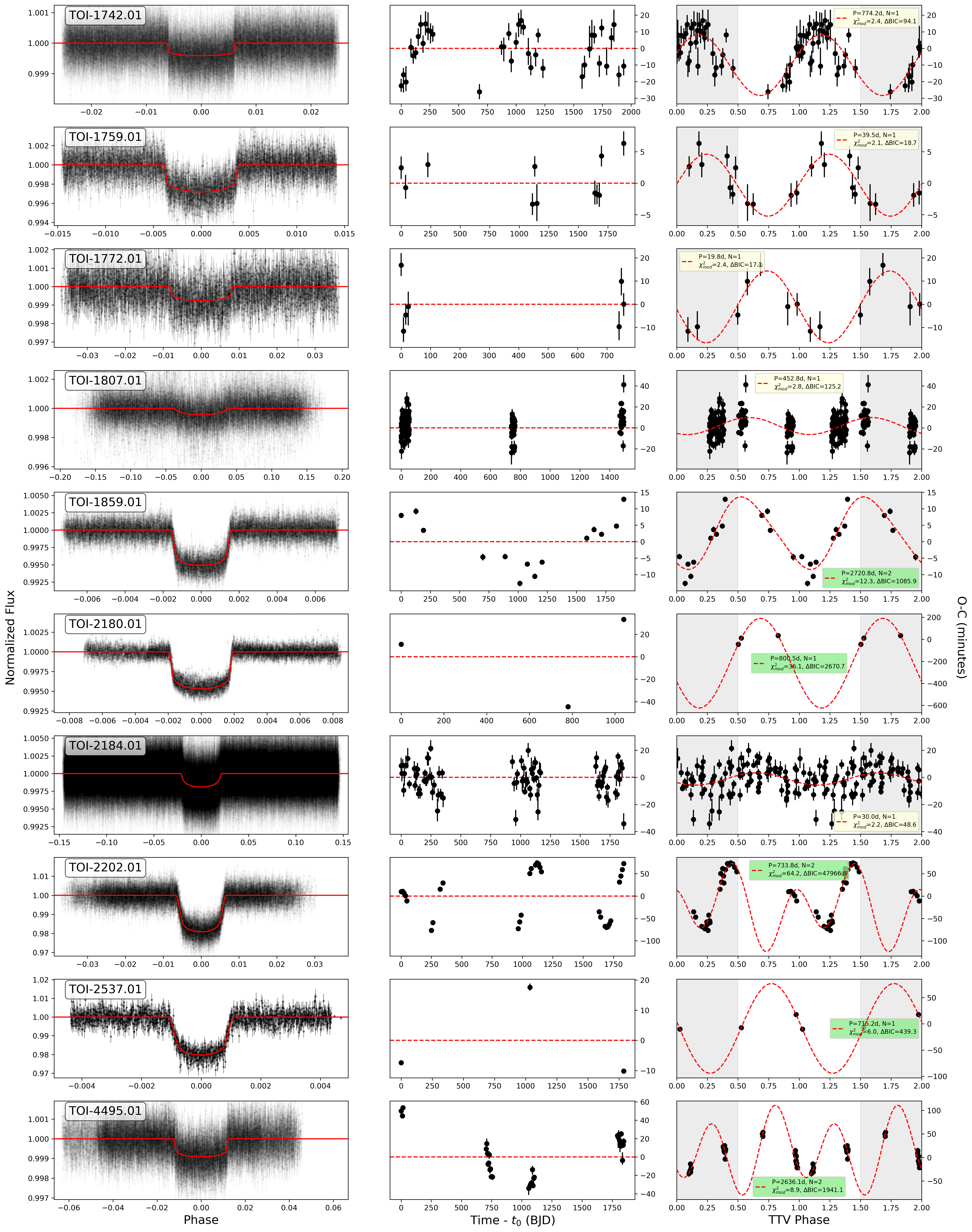}
\caption{Continuation of Fig.\,\ref{fig:TTV_1}.}
\label{fig:TTV_2}
\end{figure*}

\begin{figure*}
\centering
\includegraphics[width=1\textwidth]{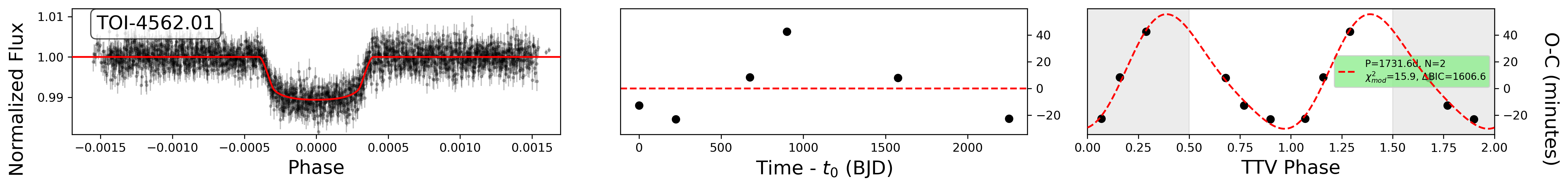}
\caption{Continuation of Fig.\,\ref{fig:TTV_2}.}
\label{fig:TTV_3}
\end{figure*}

\subsection{Strong TTVs confirmed in earlier studies}\label{sec:already_confirmed}

Among the planets that we find to have significant TTVs, there are some that have already been reported in previous studies. Nevertheless, they were included in this survey because they are still listed as single-planet systems in ExoFOP\footnote{The same is true for other systems with multiple confirmed transiting planets (e.g. TOI-1255) but no TTVs, since they were reported as single-planetary systems in ExoFOP.}. The recovery of such systems highlights the robustness of our analysis and its consistency with earlier works, including those employing different methodologies:

\begin{itemize}
    \item \textbf{TOI-199\,b}: A Saturn-type planet ($M_b\sim0.17\,M_{\mathrm{Jup}}$, $R_b\sim0.81\,R_{\mathrm{Jup}}$) orbiting a G9\,V star on a relatively long $\sim105$\,d period \citep{Hobson2023}. Its clear timing variations have been employed in the discovery paper to reveal the existence of an outer Saturn-mass planet ($M_c\sim0.28\,M_{\mathrm{Jup}}$) with an orbital period of $\sim274$\,d.
    \item \textbf{TOI-2180\,b:} A long period Jupiter ($P\sim260$\,d, $M_b\sim2.8\,M_{\mathrm{Jup}}$, $R_b\sim1.0\,R_{\mathrm{Jup}}$) orbiting a slightly evolved G5\,V star \citep{Dalba2022}. Despite being observed in only three transits, clear TTVs have been reported \citep{Dalba2024} that are possibly linked to an outer massive companion, consistent with the long-term trend seen in archival radial velocity (RV) data.
    \item \textbf{TOI-2202\,b}: A warm Jupiter ($M_b\sim0.90\,M_{\mathrm{Jup}}$, $R_b\sim0.98\,R_{\mathrm{Jup}}$) orbiting a K8\,V star every $\sim11.91$\,d \citep{Trifonov2021}. The large TTVs of this planet ($A_{\mathrm{TTV}}\sim100$\,min, $P\sim750$\,d) have been used in the discovery paper to announce the presence of another Jupiter-mass planet ($M_c\sim0.37\,M_{\mathrm{Jup}}$) in a 2:1 MMR with TOI-2202\,b. We note that the number of transits has tripled since then, and an updated N-body analysis would greatly benefit from this expanded data set.
    \item \textbf{TOI-2537\,b}: A warm Jupiter ($P\sim94$\,d, $M_b\sim1.3\,M_{\mathrm{Jup}}$, $R_b\sim1.0\,R_{\mathrm{Jup}}$) hosted by a K3\,V star \citep{Heidari2025}. With only three available transits, the TTVs of this planet are evident and have already been reported in the discovery paper as due to the interaction with an outer Jupiter-mass planet, TOI-2537\,c, first discovered through RVs.
    \item \textbf{TOI-4562\,b}: A young ($<700$\,Myr), very eccentric ($e\sim0.76$) temperate Jupiter ($M_b\sim2.30\,M_{\mathrm{Jup}}$, $R_b\sim1.12\,R_{\mathrm{Jup}}$) orbiting a F7\,V star every $\sim225$\,d \citep{Heitzmann2023}. Its large TTVs ($A_{\mathrm{TTV}}\sim43$\,min, $P\sim1700$\,d) have been linked to the presence of a second Jupiter-mass planet ($M_c\sim5.8,M_{\mathrm{Jup}}$) orbiting with a period of $\sim4000$\,d \citep{Fermiano2024}, the longest orbital period discovered to date via TTVs.
\end{itemize}

\subsection{Strong TTV candidates}\label{sec:candidates}

We further identify six new strong candidates in our sample, as defined in Sect.\,\ref{sec:classification}, that, to our knowledge, have never been announced before. In particular:

\begin{itemize}
    \item \textbf{TOI-120.01 / HD\,1397\,b}: A moderately eccentric ($e\sim0.25$) warm Jupiter ($M\sim0.41\,M_{\mathrm{Jup}}$, $R\sim1.03\,R_{\mathrm{Jup}}$) orbiting a G-type subgiant with a period of $\sim11.54$\, days \citep{Nielsen2019}. The transits of this planet display a possible modulation with $P\sim42$\,d and $A_{\mathrm{TTV}}=4.3$\,minutes, though there are only 5 valid transits and more observations are needed to confirm the variation.
    \item \textbf{TOI-139\,b}: A sub-Neptune ($R\sim2.46\,R_{\oplus}$) orbiting a K5\,V star with a period of $\sim$\,11.07 d that has been statistically validated in \citet{Mistry2023}. The planet has severe TTVs with semi-amplitude $\gtrsim11$\,hours and a long period ($\gtrsim3500$\,d).
    \item \textbf{TOI-620\,b}: A low-density, Neptune-sized planet ($R\sim3.76\,R_{\oplus}$ and $M\lesssim7\,M_{\oplus}$ at $5\sigma$ confidence) transiting an M2.5\,V star with a period of $\sim5.10$ days \citep{Refee2022}. Our analysis reveals variations with semi-amplitudes of $\sim7$\,min on a timescale of $\sim500$\,d. Given that the discovery paper reported a planet candidate with $P_{\mathrm{orb}}\sim17.7$\,d, a careful follow-up of this target is warranted to assess whether the candidate could be responsible for the observed signal.
    \item \textbf{TOI-1611.01 / HD\,207897\,b}: A sub-Neptune ($R\sim2.46\,R_{\oplus}$, $M\sim14.4\,M_{\oplus}$) orbiting a K0\,V star with a period of $\sim16.20$\,d \citep{Heidari2022}. Its transits show a variation with semi-amplitude of $\sim6.5$ min and a plausible period of $\sim900$\,d.
    \item \textbf{TOI-1859\,b}: An eccentric ($e\sim0.57$) warm Jupiter ($R\sim0.87\,R_{\mathrm{Jup}}$) orbiting a F6\,V star on a misaligned orbit with period $\sim63.48$\,d \citep{Dong2023}. The transits of this planet show a convincingly long TTV ($P\sim2700$\,d) with semi-amplitude $\sim9$\,min.
    \item \textbf{TOI-4495\,b}: A Neptune-sized planet ($R\sim3.63\,R_{\oplus}$) orbiting an evolved F-type star with a period of $\sim5.18$\,d, statistically validated in \citet{Hord2024}. The planet shows clear TTVs ($A_{\mathrm{TTV}}\sim75$\,min, $P\sim2600$\,d) which may actually be related to a non-confirmed candidate in a 2:1 mean motion orbital resonance (MMR) with TOI-4495\,b.
\end{itemize}

\subsection{Additional noteworthy systems}\label{sec:weak}

Finally, we list a few candidates that did not pass the \textit{strong} threshold criteria, though we still consider them worth mentioning:

\begin{itemize}
    \item \textbf{TOI-1227\,b}: A young ($\sim11$\,Myr) warm Jupiter ($R_b\sim0.85\,R_{\mathrm{Jup}}$) orbiting a low-mass M5\,V star with an orbital period of $\sim27.36$\,d \citep{Mann2022}. There are only 5 valid transits available, but one in particular is $\sim50$\,min late and could signal the presence of a $\sim300$\,d long TTV ($\chi^2_{\mathrm{mod}}=2.1$, $\Delta\mathrm{BIC}=56.8$). These variations have already been reported by \citealt{Almenara2024}, who suggested the presence of a 3:2 MMR $\sim6\,M_{\oplus}$ sub-Neptune. This points to the potential relevance of even our weaker candidates.
    \item \textbf{TOI-1742\,b}: A moderately eccentric ($e\sim0.3$) sub-Neptune ($M\sim9.7\,M_{\oplus}$, $R\sim2.37\,R_{\oplus}$) orbiting a solar-type star every $\sim21.27$\,d \citep{Polanski2024}. The transits show a periodic variation ($P\sim775$\,d) with a semi-amplitude of $\sim18$\,min. The large $\Delta$BIC ($\sim94$) is offset by a modest $\chi^2_{\mathrm{mod}}=2.4$, and the target is therefore classified as \textit{weak}, though it still warrants further investigation.
    \item \textbf{TOI-1807\,b}: An ultra-short period ($\sim0.55$\,d) super-Earth ($M\sim2.6\,M_{\oplus}$, $R\sim1.4\,R_{\oplus}$) orbiting a K3\,V star \citep{Nardiello2022}. This planet is barely below the threshold for \textit{strong} candidates, with $\chi^2_{\mathrm{mod}}=2.8$ and $\Delta\mathrm{BIC}=125$, though it displays convincing TTVs with $A_\mathrm{TTV}\sim8$\,min and $P\sim450$\,d. However, because the transits are quite shallow, some residual scatter remains in the O–C diagram, suggesting that more careful studies are warranted.
\end{itemize}

\subsection{Correlation analysis}

\begin{figure}
\centering
\includegraphics[width=0.5\textwidth]{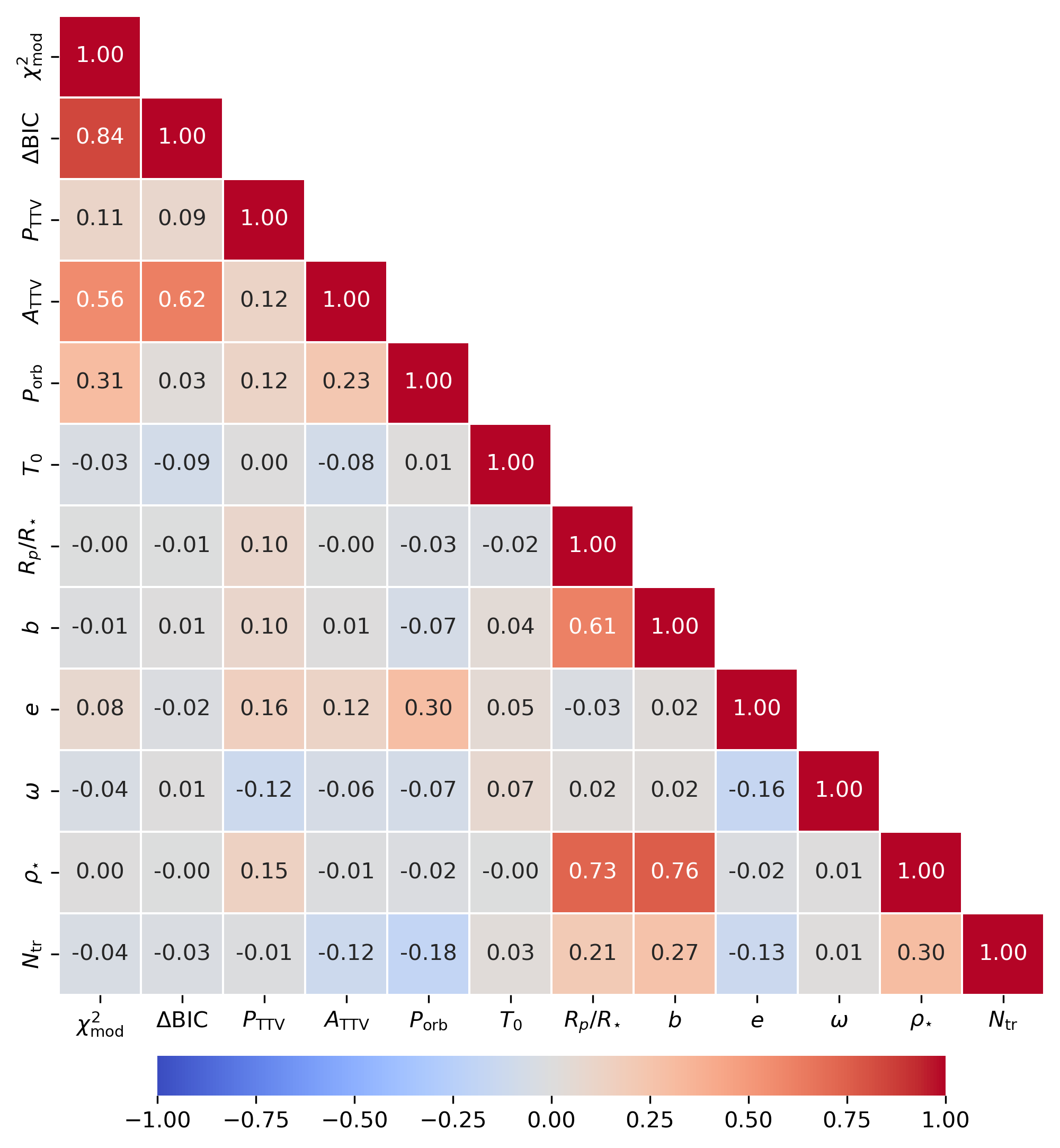}
\caption{Pearson correlation map of the parameters listed in Table\,\ref{tab:ttv_summary} for all our sample. The Pearson coefficient $r$ quantifies the strength and direction of linear relationships between pairs of variables ($r=1$ and $r=-1$ indicate perfect positive and negative correlation, while $r=0$ indicates no correlation).}
\label{fig:Pearson}
\end{figure}

We computed the Pearson correlation \citep{Pearson1895} matrix of the key physical and statistical quantities of our sample to investigate which parameters drive the detectability of TTVs (Fig.\,\ref{fig:Pearson}). The analysis shows that TTV significance is primarily driven by the amplitude $A_{\mathrm{TTV}}$, which correlates moderately with both $\chi^2_{\mathrm{mod}}$ ($r=0.40$) and $\Delta \mathrm{BIC}$ ($r=0.43$). Larger deviations thus yield more robust detections, unsurprisingly. The two significance metrics, $\chi^2_{\mathrm{mod}}$ and $\Delta \mathrm{BIC}$, are strongly correlated with each other ($r=0.84$), validating their internal consistency as indicators of the presence of a signal. 

Furthermore, the correlations among stellar density, impact parameter, and $R_p/R_\star$ reflect parameter covariances in transit modelling. Even though $\rho_\star$ is constrained by external priors, the degeneracy between $b$ and $R_p/R_\star$ propagates into the joint posterior, producing spurious trends (e.g., \citealt{Sandford2017}). In contrast, the positive correlation between eccentricity and orbital period reflects a genuine astrophysical trend that can be explained by tidal circularization: close-in planets undergo orbital damping due to stellar tides, driving them toward circular orbits, whereas planets on wider orbits remain largely unaffected and retain their primordial eccentricities \citep{Goldreich1966, Rasio1996}.

A more complex picture emerges when considering the orbital period. A weak positive trend is observed between $P_{\mathrm{orb}}$ and $\chi^2_{\mathrm{mod}}$ ($r=0.31$), but this correlation vanishes when $\Delta \mathrm{BIC}$ is considered ($r=0.03$). This likely arises because long-period planets have fewer observed transits: although individual timing deviations can be substantial, the limited number of data points reduces the statistical significance once the increased model complexity is penalized by $\Delta\mathrm{BIC}$.

Finally, neither $R_p/R_\star$ nor the eccentricity show any appreciable dependence on $\Delta \mathrm{BIC}$ ($r=-0.005$ and $r=-0.018$, respectively). Overall, the statistical detectability of TTVs appears to be dictated only by the signal amplitude and likely driven by the system architecture alone, such as proximity to MMR.

\section{Conclusions}\label{sec:conclusions}

We have conducted a systematic and homogeneous TTV survey of 423 single-planet systems discovered by the TESS mission and already confirmed or validated by follow-up studies. We do not find any clear correlation between the presence or amplitude of TTVs and the physical or orbital properties of the transiting planets. This suggests that TTVs in our sample are primarily driven by the underlying system architecture, presumably by the proximity to MMR with unseen companions.

Within the scope of this survey, our main goal was to provide a well-motivated list of candidates for detailed follow-up. To achieve this, we employed a homogeneous methodology combining precise timing measurements from a GP-based Bayesian framework with a rigorous, two-tiered statistical test for candidate classification. This quantitative framework, which requires both strong model evidence ($\Delta$BIC) and significant scatter statistic ($\chi^2_\mathrm{mod}$), ensures a high level of confidence in our strong candidates and will help increase the number of known multi-planet systems, with direct implications for demographic studies. While our analysis provides a uniform, large-scale assessment, we emphasize that individual systems may merit more detailed, case-by-case study; we therefore provide the full set of O$-$C values to allow others to explore or follow up targets of particular interest.\\

The results of this work, together with future follow-up efforts, pave the way for a deeper understanding of the architectures of planetary systems and the demographics of dynamically interacting systems discovered by TESS. The resulting catalogue of TTV candidates, along with their detailed diagnostic plots and summary parameters, offers a foundation for further work by the community. In particular, the strong candidates here presented should be considered prime targets for follow-up observations and for detailed N-body modelling to precisely determine the masses and orbits of any perturbing bodies. Looking ahead, a comprehensive TTV analysis of \textit{all} TESS candidate planets (currently about 7500), in addition to the already confirmed multi-planet systems, could further contribute to the discovery of new planets and to a refined demographic picture. Such an effort, however, lies beyond the scope of this work, as it will require a substantially larger investment of time and resources.

\begin{acknowledgements}
The author thanks the anonymous referee for their valuable comments and suggestions, which helped improve the quality of this manuscript. The author is also grateful to the colleagues at INAF - Osservatorio Astrofisico di Torino for helpful and stimulating conversations related to the TTV analysis. The author further acknowledges financial contribution from the INAF Large Grant 2023 ``EXODEMO''. The present work has made extensive use of the 48-core HOT-ATMOS server at INAF - Osservatorio Astrofisico di Torino. This paper uses data from the TESS mission, whose funding is provided by the NASA Science Mission directorate. We acknowledge the use of the TESS archive, which is supported by the NASA Exoplanet Archive, and the ExoFOP-TESS service, which are operated by the California Institute of Technology, under contract with NASA. This work has made use of the Python packages \texttt{numpy}, \texttt{pandas}, \texttt{matplotlib}, \texttt{lightkurve}, \texttt{dynesty}, \texttt{batman}, \texttt{astroquery}, \texttt{juliet} and packages therein.
\end{acknowledgements}

\bibliographystyle{aa}
\bibliography{TTV}

\appendix

\onecolumn
\section{Additional table}


\setlength{\tabcolsep}{1pt} 
{\scriptsize 
\begin{longtable}{lrrrrrrrrrrrrr}
\caption{Summary of the TTV analysis results.} \label{tab:ttv_summary} \\
\toprule
TOI & TTVs & $\chi^2_{\mathrm{mod}}$ & $\Delta$BIC & $P_{\mathrm{TTV}}$ & $A_{\mathrm{TTV}}$ & $P_{\mathrm{orb}}$ & $T_0$ & $R_p/R_{\star}$ & $b$ & $e$ & $\omega$ & $\rho_{\star}$ & $N_{\mathrm{tr}}\,(^\dagger)$ \\
& & & & [d] & [min] & [d] & [TBJD] & & & & & [cgs] & \\
\midrule
\endfirsthead
\caption[]{-- continued from previous page} \\
\toprule
TOI & TTVs & $\chi^2_{\mathrm{mod}}$ & $\Delta$BIC & $P_{\mathrm{TTV}}$ & $A_{\mathrm{TTV}}$ & $P_{\mathrm{orb}}$ & $T_0$ & $R_p/R_{\star}$ & $b$ & $e$ & $\omega$ & $\rho_{\star}$ & $N_{\mathrm{tr}}\,(^\dagger)$ \\
& & & & [d] & [min] & [d] & [TBJD] & & & & & [cgs] & \\
\midrule
\endhead
\bottomrule
\endfoot
\bottomrule
\endlastfoot
118.01 & No & 1.9 & 8.6 & 19.1 & 2.7 & $6.0344703^{+0.0000031}_{-0.0000034}$ & $3175.74963^{+0.00065}_{-0.00067}$ & $0.04203^{+0.00053}_{-0.00052}$ & $0.8390^{+0.0092}_{-0.0090}$ & 0 & 90 & $1.205^{+0.089}_{-0.083}$ & 12 \\
120.01 & Yes & \textbf{4.1} & \textbf{33.6} & 41.9 & 3.8 & $11.53622^{+0.00021}_{-0.00026}$ & $1332.08351^{+0.00051}_{-0.00049}$ & $0.04501^{+0.00019}_{-0.00020}$ & $0.18^{+0.08}_{-0.10}$ & $0.252^{+0.020}_{-0.019}$ & $3.5^{+4.8}_{-4.5}$ & $0.1482^{+0.0078}_{-0.0080}$ & 5 \\
122.01 & No & 0.5 & -0.0 & 15.1 & 4.2 & $5.0780375^{+0.0000072}_{-0.0000072}$ & $1329.1188^{+0.0018}_{-0.0019}$ & $0.0760^{+0.0021}_{-0.0022}$ & $0.654^{+0.067}_{-0.074}$ & 0 & 90 & $15.1^{+4.2}_{-3.6}$ & 20 \\
123.01 & No & 1.0 & 11.8 & 7.9 & 0.8 & $3.30887883^{+0.00000034}_{-0.00000033}$ & $3175.03799^{+0.00011}_{-0.00011}$ & $0.06022^{+0.00021}_{-0.00022}$ & $0.442^{+0.024}_{-0.025}$ & 0 & 90 & $0.1444^{+0.0052}_{-0.0053}$ & 27 \\
127.01 & No & 0.9 & 7.8 & 23.1 & 1.2 & $3.79551764^{+0.00000095}_{-0.00000095}$ & $1325.65055^{+0.00028}_{-0.00028}$ & $0.2178^{+0.0013}_{-0.0015}$ & $0.365^{+0.022}_{-0.022}$ & 0 & 90 & $6.24^{+0.15}_{-0.14}$ & 36 \\
128.01 & No & 1.0 & 19.6 & 27.3 & 2.7 & $4.9404764^{+0.0000025}_{-0.0000036}$ & $3175.86830^{+0.00050}_{-0.00054}$ & $0.02061^{+0.00052}_{-0.00036}$ & $0.663^{+0.083}_{-0.078}$ & 0 & 90 & $1.80^{+0.49}_{-0.53}$ & 78 \\
129.01 & No & 1.0 & 6.1 & 3.1 & 0.2 & $0.980972180^{+0.000000048}_{-0.000000048}$ & $3205.646923^{+0.000066}_{-0.000065}$ & $0.282^{+0.049}_{-0.046}$ & $1.161^{+0.055}_{-0.052}$ & 0 & 90 & $2.887^{+0.068}_{-0.067}$ & 128 \\
130.01 & No & 0.6 & 2.3 & 250.4 & 4.0 & $14.339186^{+0.000026}_{-0.000026}$ & $1336.5585^{+0.0014}_{-0.0014}$ & $0.01876^{+0.00022}_{-0.00023}$ & $0.28^{+0.14}_{-0.16}$ & 0 & 90 & $1.11^{+0.12}_{-0.17}$ & 12 \\
\end{longtable}
\tablefoot{$(^\dagger)$ $N_{\mathrm{tr}}$ is the number of transits analyzed per planet. The O$-$C list for each planet is available in the digital version of the paper.}
}
\normalsize




\end{document}